\documentclass[twocolumn,aps,prd]{revtex4-2}
\usepackage{graphicx,amssymb}
\usepackage[utf8]{inputenc} % Required for inputting international characters
 
\usepackage[urlcolor=blue,colorlinks,breaklinks=true]{hyperref}

\PassOptionsToPackage{obeyspaces,spaces,hyphens}{url}
 
\usepackage{amsmath}

\begin{document}
 	%My commands
 	\def\half{{1\over2}}
 	\def\shalf{\textstyle{{1\over2}}}
 	
 	\newcommand\lsim{\mathrel{\rlap{\lower4pt\hbox{\hskip1pt$\sim$}}
 			\raise1pt\hbox{$<$}}}
 	\newcommand\gsim{\mathrel{\rlap{\lower4pt\hbox{\hskip1pt$\sim$}}
 			\raise1pt\hbox{$>$}}}

\newcommand{\be}{\begin{equation}}
\newcommand{\ee}{\end{equation}}
\newcommand{\bq}{\begin{eqnarray}}
\newcommand{\eq}{\end{eqnarray}}
 	
%\title{On-shell Lagrangian of particles and fluids: energy-momentum constraints in theories of gravity nonminimally coupled to matter}

\title{On-shell Lagrangian of an ideal gas}
 	 	
\author{P.P. Avelino}
\email[Electronic address: ]{pedro.avelino@astro.up.pt}
\affiliation{Departamento de F\'{\i}sica e Astronomia, Faculdade de Ci\^encias, Universidade do Porto, Rua do Campo Alegre 687, PT4169-007 Porto, Portugal}
\affiliation{Instituto de Astrof\'{\i}sica e Ci\^encias do Espa{\c c}o, Universidade do Porto, CAUP, Rua das Estrelas, PT4150-762 Porto, Portugal}

\author{R.P.L. Azevedo}
\email[Electronic address: ]{rplazevedo@astro.up.pt}
\affiliation{Departamento de F\'{\i}sica e Astronomia, Faculdade de Ci\^encias, Universidade do Porto, Rua do Campo Alegre 687, PT4169-007 Porto, Portugal}
\affiliation{Instituto de Astrof\'{\i}sica e Ci\^encias do Espa{\c c}o, Universidade do Porto, CAUP, Rua das Estrelas, PT4150-762 Porto, Portugal}

\date{\today}

\date{\today}
\begin{abstract}
In the context of general relativity, both energy and linear momentum constraints lead to the same equation for the evolution of the speed of free localized particles with fixed proper mass and structure in a homogeneous and isotropic Friedmann-Lemaître-Robertson-Walker universe. In this paper we extend this result by considering the dynamics of particles and fluids in the context of theories of gravity nonminimally coupled to matter. We show that the equation for the evolution of the linear momentum of the particles may be obtained irrespective of any prior assumptions regarding the form of the on-shell Lagrangian of the matter fields. We also find that consistency between the evolution of the energy and linear momentum of the particles requires that their volume-averaged on-shell Lagrangian and energy-momentum tensor trace coincide ($\mathcal L_{\rm on-shell}=T$). We further demonstrate that the same applies to an ideal gas composed of many such particles. This result implies that the two most common assumptions in the literature for the on-shell Lagrangian of a perfect fluid ($\mathcal L_{\rm on-shell}=\mathcal{P}$ and $\mathcal L_{\rm on-shell}=-\rho$, where $\rho$ and $\mathcal{P}$ are the proper density and pressure of the fluid, respectively) do not apply to an ideal gas, except in the case of dust (in which case $T=-\rho$).
\end{abstract}

\maketitle
 	
\section{Introduction}
\label{sec:intr}

The energy-momentum content of the Universe is often described macroscopically as a collection of minimally coupled fluids --- usually perfect fluids with no shear stresses, viscosity or heat conduction --- without an explicit reference to the Lagrangians which describe their microscopic dynamics. In the context of general relativity this is not a problem, since the matter Lagrangians do not enter explicitly in the equations of motion of the gravitational and matter fields \cite{1970PhRvD...2.2762S, Schutz1977,  Brown1993a}. 
However,  this is no longer the case in theories in theories of gravity with a nonminimal coupling (NMC) between matter and curvature. As a result of such couplings, the energy-momentum tensor is not, in general, covariantly conserved and the on-shell matter Lagrangians can directly affect the dynamics of the gravitational and matter fields (see, for example, \cite{Bertolami2007, Bertolami:2008ab, Harko:2011kv, 2018EPJC...78..326B, 2018PhRvD..97j4041B, 2018PhRvD..98l4020M, 2019PhRvD..99l4027F}). Hence, caution must be taken when assuming a specific form for the on-shell Lagrangian. In fact, the various on-shell Lagrangians that have been used in the literature to describe perfect fluids (such as $\mathcal{L}_\text{on-shell}= -\rho$ \cite{Brown1993a,Harko2010a,Minazzoli2012,Minazzoli2013,Ferreira:2020fma,Arruga2021}, $\mathcal{L}_\text{on-shell}=\mathcal{P}$ \cite{Brown1993a,Ferreira:2020fma} or $\mathcal{L}_\text{on-shell}=T$ \cite{Avelino:2018rsb,Avelino:2018qgt,Ferreira:2020fma}) almost always lead to different physical predictions in the context of theories of gravity with a NMC between matter and curvature. Therefore, these Lagrangians cannot generally be used to describe the same fluid, thus implying that prior knowledge of their appropriate form may be crucial for a correct characterization of the overall dynamics. This is also true in the presence of an NMC between different physical matter components even in the absence of an NMC to gravity \cite{Bekenstein1982,Sandvik2002,Anchordoqui2003,Copeland2004,Lee2004,Koivisto2005,Avelino2008,Bettoni2011,Ayaita2012,Pourtsidou2013,Bohmer2015,Bohmer2015a,Bettoni2015,Koivisto2015,Brax2016,Tamanini2016,Dutta2017,Barros2019,Kase2020a}.

There is no universal on-shell Lagrangian of a fluid (see \cite{Ferreira:2020fma} for a recent discussion of the Lagrangian description of cosmic fluids). Even in the case of a perfect fluid, the  on-shell Lagrangian depends in general on its microscopic properties. In fact, it is not hard to find examples of perfect fluids which can have different on-shell Lagrangians but the same energy-momentum tensor. Still, it has been shown that the Lagrangian of any fluid which can be approximated as a collection of moving point particles of fixed mass, whose motion might be subject to multiple point-like collisions, is given by the trace of its energy-momentum tensor \cite{Avelino:2018rsb,Avelino:2018qgt} ($\mathcal L_{\rm on-shell}=T$). This ideal gas approximation provides a good description of a significant part of the energy content of the Universe, including dark matter, baryons and photons, but cannot be used to describe dark energy \cite{Ferreira:2020fma} or a perfect fluid whose off-shell Lagrangian depends solely on the particle number density \cite{Harko2010a,Minazzoli2012,Minazzoli2013,Ferreira:2020fma}.

The non-minimal coupling between gravity and the matter fields generally gives rise to additional forces dependent on the individual linear momentum of the particles. These forces may have a cosmological impact, in particular on the cosmic microwave background and primordial nucleosynthesis \cite{Avelino:2018rsb,Azevedo:2018nvi}, but can also play a non-negligible role on microscopic scales \cite{Fisher:2021nwm}. They might also potentially lead to violations of Etherington's distance-duality relation \cite{Azevedo2021}, of Boltzmann's H-theorem \cite{Avelino:2020fek} and of the second law of thermodynamics \cite{Azevedo:2019oah,Avelino:2020fek}. In this paper we revisit the dynamics of localized particles of fixed mass and structure, and of ideal gases made up of such particles in the light of the necessary consistency between the energy and the momentum constraints on their dynamics --- we shall make no specific assumptions regarding the particles' composition. Particular attention will be devoted to the constraints on the appropriate form of the on-shell Lagrangians. 

The outline of this paper is as follows. In Sec. \ref{sec:emGR} we briefly describe how the same evolution equation for the speed of free localized particles of fixed mass and structure in a homogeneous and isotropic Friedmann-Lemaître-Robertson-Walker (FLRW) universe can be obtained in five different ways by taking into account energy or linear momentum conservation. In Sec. \ref{sec:emNMC} we extend the analysis of Sec. \ref{sec:emGR} by considering the dynamics of particles and fluids in FLRW universes in the context of theories of gravity nonminimally coupled to matter. We start by deriving the evolution of the particles' linear momentum independently of any prior assumptions regarding the form of the on-shell Lagrangian of the matter fields.  We then demonstrate that the necessary consistency between energy and linear momentum evolution, both of the individual particles and of ideal gases composed of many such particles, uniquely defines the appropriate form of the corresponding on-shell Lagrangians. Finally, we conclude in section \ref{sec:conc}.

Throughout this paper we use units such that $c=16 \pi G =1$, where $c$ is the value of the speed of light in vacuum, and $G$ is the gravitational constant.  We also adopt the metric signature $(-,+,+,+)$. The Einstein summation convention will be used when a Greek or Latin index appears twice in a single term, once in an upper (superscript) and once in a lower (subscript) position. Greek and Latin indices take the values $0,\dots, 3$ and $1,\dots, 3$, respectively.

\section{Energy-momentum conservation and the dynamics of particles and fluids}
\label{sec:emGR}

In this section we shall present five different derivations of the equation for the evolution of the speed of free localized particles of fixed mass and structure in a homogeneous and isotropic FLRW universe, relying solely on linear momentum and energy conservation. 

\subsection{Energy-momentum conservation in General Relativity}

Let us start by considering the Einstein-Hilbert action
\be
S=\int (R+{\mathcal L}_{\rm m}) {\sqrt {-g}}d^4 x \,,
\ee
where ${\mathcal L}_{\rm m}$ is the matter Lagrangian, $R$ is the Ricci scalar, $g=\det (g_{\mu\nu})$ and $g_{\mu\nu}$ are the components of the metric tensor. In general relativity the energy-momentum tensor of the matter fields, whose components are given by
\be
T^{\mu\nu}=-\frac{2}{{\sqrt {-g}}} \frac{\delta({\mathcal L}_{\rm m}{\sqrt {-g}})}{\delta g_{\mu \nu}}=-2\frac{\delta{\mathcal L}_{\rm m}}{\delta g_{\mu \nu}}+g^{\mu\nu} {\mathcal L}_{\rm m} \label{T}\,,
\ee
is covariantly conserved, so that
\be
\nabla_{\nu} {T{^\nu}}_\mu= 0\,. \label{emconservation}
\ee
Throughout this paper we shall consider either the energy-momentum tensor of the individual particles with components $T^{\mu \nu}$ or the energy-momentum tensor of a perfect fluid composed of many such particles. The components of the latter are given by
\begin{equation}
{\mathcal T}^{\mu\nu}=(\rho+\mathcal P)U^\mu U^\nu + \mathcal P g^{\mu\nu}\label{EMPF}\,,
\end{equation}
where $\rho$, $~\mathcal P$ and $U^\mu$ are respectively the proper energy density, the proper pressure and the components of the 4-velocity of the perfect fluid --- notice the use of a different letter to identify the energy-momentum tensor of a perfect fluid. Energy-momentum conservation implies that
\begin{equation}
h^{\mu \beta} \nabla_{\alpha} {{\mathcal T}{^\alpha}}_\beta = (\rho+\mathcal P)U^\nu  \nabla_\nu U^\mu  + h^{\mu \beta}  \nabla_\beta \mathcal P= 0 \,,
\end{equation}
where $h^{\mu \nu}=g^{\mu \nu}+ U^\mu U^\nu$ is the projection operator. In the case of dust $\mathcal{P}=0$ and, therefore, 
\begin{equation}
U^\nu  \nabla_\nu U^\mu  = 0 \,.
\end{equation}

\subsection{Energy and momentum of particles in a Minkowsky spacetime}

Consider a single particle and a rest frame where its energy-momentum tensor is static. Assuming that the gravitational interaction plays a negligible role on the particle structure,  the spacetime in and around the particle may be described by a Minkowski metric line element
\begin{equation}
ds^2=-dt^2 +d \vec r \cdot d \vec r =-dt^2  +dx^2+dy^2+dz^2 \,,
\end{equation}
where $t$ is the physical time and $\vec r = (x,y,z)$ are cartesian coordinates.

The particle's proper frame is defined by
\begin{equation}
\int {T^i}_{0[\rm prop]} \, d^3 r_{[\rm prop]}= -\int {T^0}_{i[\rm prop]} \, d^3 r_{[\rm prop]} =0 \,, \label{Ti0}
\end{equation}
with 
\begin{equation}
E_{[\rm prop]}  = - \int {T^0}_{0[\rm prop]} \, d^3 r  \label{Eprop}
\end{equation}
being the proper energy of the particle (the subscript $[\rm prop]$ is used to designate quantities evaluated in the proper frame). On the other hand, the generalized von Laue conditions \cite{doi:10.1002/andp.19113400808,Avelino:2018qgt},
\begin{eqnarray}
\int {T^1}_{1[\rm prop]}  \, d^3 r_{[\rm prop]}  &=& \int {T^2}_{2[\rm prop]}  \, d^3 r_{[\rm prop]}  \nonumber \\
&=&  \int {T^3}_{3[\rm prop]}  \, d^3 r _{[\rm prop]} = 0 \,, \label{vonlaue}
\end{eqnarray}
are required for particle stability.

Consider a Lorentz boost in the $x$ direction defined by
\begin{eqnarray}
t&=&\gamma(t_{[\rm prop]}+vx_{[\rm prop]})\,,\\
x&=&\gamma(x_{[\rm prop]}+vt_{[\rm prop]})\,,\\
y&=&y_{[\rm prop]}\,,\\
z&=&z_{[\rm prop]}\,,
\end{eqnarray}
where $\gamma=\left(1-v^{2}\right)^{-1/2}$ is the Lorentz factor and $v$ is the particle velocity. Under this boost, the components of the energy-momentum tensor $T_{\mu\nu}$ transform
as
\begin{equation}
{T^{\mu}}_{\nu}={\Lambda^{\mu}}_{\alpha}\,{\Lambda_{\nu}}^{\beta}\,{T^{\alpha}}_{\beta[\rm prop]}\label{emtransf}
\end{equation}
where the non-zero components of ${\Lambda^\mu}_\alpha$ and ${\Lambda_\nu}^\beta$ are
\begin{eqnarray}
{\Lambda^0}_0&=&{\Lambda^1}_1={\Lambda_0}^0={\Lambda_1}^1=\gamma\,,\label{lorentz1}\\
{\Lambda^0}_1&=&{\Lambda^1}_0=-{\Lambda_0}^1=-{\Lambda_1}^0=\gamma v\,,\label{lorentz2}\\
{\Lambda^2}_2&=&{\Lambda^3}_3={\Lambda_2}^2={\Lambda_3}^3=1\label{lorentz3}\,,
\end{eqnarray}
with all other components vanishing. In the moving frame the energy and linear momentum of the particle are given, respectively, by
\begin{eqnarray}
E &=& - \int {T^0}_0 \, d^3 r =E_{[\rm prop]}  \gamma  \,, \label{Eeq}\\
p &=& \int {T^1}_0 \, d^3 r=E_{[\rm prop]} \gamma v =E v \label{peq}\,,
\end{eqnarray}
where Eqs. (\ref{Ti0}),  (\ref{Eprop}), (\ref{emtransf}),  (\ref{lorentz1}), (\ref{lorentz2}), as well as Lorentz contraction, have been taken into account in the derivation of Eqs. (\ref{Eeq}) and  (\ref{peq}). These two  equations imply that $E^2-p^2=E_{\rm [prop]}^2$ and
\begin{equation}
\dot p  = \dot E  \frac{E}{p}=  \frac{\dot E}{v}=E_{[\rm prop]} \dot v \gamma^3\label{dotp}\,.
\end{equation}

On the other hand, using Eqs.  (\ref{emtransf}), (\ref{lorentz2}) and  (\ref{lorentz3}) one finds
\begin{eqnarray}
\int {T^1}_1 \, d^3 r&=& E_{[\rm prop]}  \gamma v^2 = E v^2\,,\label{T11}\\
\int {T^2}_2 \, d^3 r &=& \int {T^3}_3 \, d^3 r = 0  \,,
\end{eqnarray}
so that
\begin{equation}
\int {T^i}_i\, d^3 r=E_{[\rm prop]}\gamma v^2 = E v^2\,. \label{trace}
\end{equation}
Also notice that
\begin{equation}
\int T \, d^3 r= \int {T^\mu}_\mu\, d^3 r=-\frac{E_{[\rm prop]}}{\gamma} = -\frac{E}{\gamma^2}\,. \label{traceT}
\end{equation}

\subsection{Free particles in an FLRW spacetime}

In a flat homogeneous and isotropic universe, described by the FLRW metric, the line element may be written as
\begin{equation}
ds^2=a^2(\eta)(-d\eta^2+d \vec q \cdot d \vec q\,) \,,
\end{equation}
where $a(\eta)$ is the scale factor, $\eta = \int dt/a$ is the conformal time and $\vec q$ are comoving cartesian coordinates. In an FLRW spacetime the nonvanishing components of the connection are given by 
\begin{equation}
\Gamma_{0 0}^0={H}\,, \quad \Gamma_{i j}^0={H} \,\delta_{ij}\,, \quad \Gamma_{0 j}^i={H} \,{\delta^i}_j \,,
\end{equation}
where $H = \dot a /a$ and a dot denotes a derivative with respect to the conformal time.

\subsubsection{Linear momentum conservation}

Consider again a single free particle moving along the $x$-direction. The $x$-component of Eq. (\ref{emconservation}) describing momentum conservation in an FLRW spacetime then implies that
\begin{equation}
0=\nabla_\nu {T^\nu}_1   =  \partial_0 {T^0}_1 + \partial_i {T^i}_1  + 4  {H} {T^0}_1\,.
\end{equation}
Integrating over the spatial volume one finds that
\begin{equation}
{\dot p}+ {H} p=0\,, \label{pev}
\end{equation}
where
\begin{equation}
p=\int {T^0}_1 \, d^3 r=a^3 \int {T^0}_1 \, d^3 q \,.
\end{equation}
In this derivation we have assumed that the particle is isolated so that the energy-momentum tensor vanishes outside it. Hence,
\begin{equation}
\int \partial_i {T^\mu}_\nu \, d^3 q =0 \label{isolated}
\end{equation}
for any possible value of $\mu$, $\nu$ and $i$. Notice that Eq. (\ref{pev}) implies that $p = E_{[\rm prop]} \gamma v \propto a^{-1}$. Dividing Eq. (\ref{pev})  by $E_{[\rm prop]} $, taking into account Eq. (\ref{dotp}),  one obtains the  equation for the evolution of the free particle velocity in a homogeneous and isotropic FLRW universe:
\begin{equation}
{\dot v}+ {H}(1-v^2) v=0\,. \label{vev}
\end{equation}

\subsubsection{Energy conservation}

Energy conservation, on the other hand, implies that
\begin{equation}
0=\nabla_\nu {T^\nu}_0   =  \partial_0 {T^0}_0 + \partial_i {T^i}_0  +  3  {H} {T^0}_0 -  {H} {T^i}_i \,.
\end{equation}
Integrating over the spatial volume, and using Eqs. (\ref{trace}) and (\ref{isolated}), one finds that
\begin{equation}
{\dot E}+ {H} v^2 E=0\,, \label{Eev}
\end{equation}
where
\begin{equation}
E=-\int {T^0}_0 \,d^3 r=-a^3 \int {T^0}_0 \,d^3 q \,.
\end{equation}
Dividing Eq. (\ref{Eev}) by $v$, taking into account Eq. (\ref{dotp}), once again one obtains Eq. (\ref{pev}) for the evolution of linear momentum in a homogeneous and isotropic FLRW universe.

\subsection{Perfect fluids in an FLRW spacetime}

We shall now derive the dynamics of free particles assuming that they are part of a homogeneous perfect fluid (see Eq. (\ref{EMPF})) with the proper energy density $\rho$ and the proper pressure $\mathcal{P}$ depending only on time.

\subsubsection{Linear momentum conservation: dust}

In the case of a perfect fluid with vanishing proper pressure $\mathcal P$, the components of the energy-momentum tensor are
\begin{equation}
\label{eq:pfemt}
{\mathcal T}^{\mu\nu}=\rho \, U^\mu U^\nu \,,
\end{equation}
If the fluid moves in the positive $x$-direction, then
\begin{eqnarray}
U^0&=&\frac{d\eta}{d\tau} = \frac{\gamma}{a}\,,\\
U^1&=&\frac{dq^1}{d\tau} = {\dot q}^1\frac{d\eta}{d\tau}=v \, U^0 =v\frac{\gamma}{a} \,,\\
U^2&=&U^3=0\,.
\end{eqnarray}
The $x$-component of Eq. (\ref{emconservation}), describing momentum conservation, implies that
\be
\dot U^1 U^0+ 2 \Gamma^1_{1 0} U^0 U^1=0 \,.
\ee
Multiplying this equation by $E_{[\rm prop]} a/U^0$, taking into account that $U^1=\gamma v/a$ and that $\Gamma^0_{1 1}=H$, one obtains once again Eq. (\ref{pev}) for the evolution of linear momentum of a free particle in a homogeneous and isotropic FLRW universe. 

\subsubsection{Energy conservation: dust}

The time component of Eq. (\ref{emconservation}), describing energy conservation, is given by
\be
\dot U^0 U^0+  \Gamma^0_{0 0} U^0 U^0 + \Gamma^0_{1 1} U^1 U^1=0 \,.
\ee
Multiplying this equation by $E_{[\rm prop]} a/U^0$, taking into account that $U^0=\gamma/a$,  $U^1=\gamma v/a$, and that $\Gamma^0_{0 0}=\Gamma^0_{1 1}=H$, one obtains once again Eq. (\ref{Eev}) for the evolution of the energy of a free particle in a homogeneous and isotropic FLRW universe, which has been shown to be equivalent to Eq. (\ref{pev}) for the evolution of the linear momentum. 

\subsubsection{Energy conservation: homogeneous and isotropic fluid}
\label{sec:emGRF}

We shall now consider a homogeneous and isotropic perfect fluid (at rest in the comoving frame, so that $U^i=0$)  made up of free particles all with the same speed $v$. This fluid can be pictured as the combination of six equal density dust fluid components moving in the positive/negative $x$, $y$, and $z$ directions. The time component of Eq. (\ref{emconservation}), describing energy conservation, implies that
\begin{equation}
{\dot \rho} +  3 H (\rho +\mathcal P) =0\,. \label{rhoEQ}
\end{equation}
If the number $N$ of particles in a volume $V=a^3$ is conserved then
\be
\rho=\frac{NE}{V} = \frac{N E}{a^3}=N E_{[\rm prop]}  \frac{\gamma}{a^3}\propto \frac{\gamma}{a^3}\,. \label{rhoEV}
\ee
On the other hand, if the perfect fluid is an ideal gas then its proper pressure is given by 
\be
\mathcal P=\rho   v^2/3\,. \label{Pideal}
\ee
Substituting the conditions given in Eqs. (\ref{rhoEV}) and (\ref{Pideal}) into Eq. (\ref{rhoEQ}) multiplied by $a/N$, one again arrives at Eq. (\ref{Eev}), the same as the one derived considering energy conservation for individual free particles.

\section{Energy-momentum evolution and the dynamics of particles and fluids in NMC gravity}
\label{sec:emNMC}

In this section we shall again present five different derivations of the equation for the evolution of the speed of individual localized particles of fixed mass and structure in a homogeneous and isotropic FLRW universe, but now considering the possibility of a coupling between gravity and the matter fields. We shall demonstrate that consistency between the results obtained uniquely defines the correct form of the corresponding on-shell Lagrangians. 

\subsection{Energy-momentum constraints in NMC gravity}

Let us start by considering the action \cite{Harko:2011kv}
\be
S=\int f(R,{\mathcal L}_{\rm m}) {\sqrt {-g}}d^4 x \,, \label{action}
\ee
allowing for a NMC between gravity and the matter matter fields. In this and other NMC theories the energy momentum tensor of the matter fields, whose components are given in Eq. (\ref{T}), is not in general covariantly conserved. Instead one has
\be
\nabla_{\nu} {T_{\mu}}^\nu=  S_{\mu}\,, \label{Tncons}
\ee
where
\bq
S_{\mu} &=& ({\mathcal L}_{\rm m} {\delta^{\nu}}_\mu -{T^{\nu}}_\mu) \times \nonumber \\
&\times& \left( [\ln \left| f_{,{\mathcal L}_{\rm m}}\right|]_{,R} \nabla_\nu R + [\ln \left|f_{,{\mathcal L}_{\rm m}}\right|]_{,{\mathcal L}_{\rm m}} \nabla_\nu {\mathcal L}_{\rm m}  \right)\,.\label{Tncons1}
\eq
This implies that the knowledge of the Lagrangian of the matter fields is in general required in order to determine the  corresponding dynamics, even when considering a perfect fluid. For the sake of definiteness, let us consider the case where \cite{Bertolami2007}
\be
f(R,{\mathcal L}_{\rm m})=f_1(R) + {\mathcal L}_{\rm m} f_2(R)\,,
\ee
so that
\be
S_{\mu} = ({\mathcal L}_{\rm m} \delta_{\mu}^\nu -{T_{\mu}}^\nu) \frac{\nabla_\nu f_2}{f_2} \,.\label{Tncons2}
\ee

Here, we shall again consider either the energy-momentum tensor of the individual particles with components $T^{\mu \nu}$ or the energy-momentum tensor of a perfect fluid composed of many such particles whose components given in Eq. (\ref{EMPF}).

In the case of a perfect fluid Eq. (\ref{Tncons}), with $S_\mu$ given by Eq. (\ref{Tncons2}), implies that \cite{Bertolami2007}
\begin{equation}
U^\nu  \nabla_\nu U^\mu  = \frac{1}{\rho+\mathcal P} \left[\left(\mathcal{L}_{\rm f} -\mathcal P\right) \frac{\nabla_\nu f_2}{f_2} -\nabla_\nu \mathcal P\right]h^{\mu \nu}\,, \label{UevNMC}
\end{equation}
where ${\mathcal L}_{\rm f}$ and $h^{\mu \nu}=g^{\mu \nu}+ U^\mu U^\nu$ are the on-shell  Lagrangian of the perfect fluid and the projection operator, respectively. In the following we shall also consider the particular case of dust with $\mathcal{L}_\text{f}=\mathcal{L}_\text{dust}$ (characterized by $\mathcal{P}_{\rm dust}=0$ and $\rho_{\rm dust}=-\mathcal T_{\rm dust}$), for which
\begin{equation}
U^\nu  \nabla_\nu U^\mu  = \frac{\mathcal{L}_{\rm dust} }{\mathcal \rho_{\rm dust}} \frac{\nabla_\nu f_2}{f_2} h^{\mu \nu}\,, \label{UevNMCdust}
\end{equation}

\subsection{Free particles in FLRW spacetimes}

Consider once again the motion of localized particles of fixed mass and structure in an FLRW background, but this time in the context of NMC gravity. Given that the energy-momentum tensor is no longer covariantly constrained, the presence of additional dynamical terms, dependent on the matter Lagrangian, will need to be taken into account.

\subsubsection{Linear momentum evolution}

For a single isolated particle moving along the $x$-direction in an FLRW background, the $x$-component of Eq. (\ref{Tncons2}) implies that
\be
\int S_1 \, d^3 r =  -p   \frac{\dot f_2}{f_2} \,, \label{S1int}
\ee
Hence, considering the $x$-component of Eq. (\ref{Tncons}), and following the same steps of the previous section, the equation for the evolution of the linear momentum of the particle can now be generalized to 
\begin{equation}
{\dot p}+ \Theta  \, p=0\,, \label{pev1}
\end{equation}
where $\Theta$ is defined by
\begin{equation}
\Theta = \frac{\dot b}{b}= \frac{\dot a}{a}+ \frac{\dot  f_2}{f_2} = H + \frac{\dot  f_2}{f_2} \label{pev1theta}
\end{equation}
and $b=a f_2$. Notice that Eq. (\ref{pev1})  was obtained without making any prior assumptions about the specific form of the on-shell Lagrangian.

\subsubsection{Energy evolution}

Of course, one must be able to arrive at the same result using the time component of Eq. (\ref{Tncons}) --- otherwise there would be an inconsistency. The time component of Eq. (\ref{Tncons2}) requires that
\bq
\int S_0 \, d^3 r =  \left(\int {\mathcal L}_{\rm m} \,  d^3 r +E\right) \frac{\dot  f_2}{f_2} \,, \label{S0int}
\eq
Following the same steps of the previous section but now using the time component of Eq. (\ref{Tncons}) and Eq. (\ref{S0int}) one obtains
\begin{equation}
{\dot E}+ {H}  v^2 E=-\left(\int {\mathcal L}_{\rm m} \,  d^3 r +E\right) \frac{\dot  f_2}{f_2} \,. \label{Eev1}
\end{equation}
Dividing Eq. (\ref{Eev1}) by $v$, taking into account Eqs. (\ref{peq}) and (\ref{dotp}), one finds that 
\begin{equation}
{\dot p}+ {H}  p=-\frac{\int {\mathcal L}_{\rm m} \,  d^3 r +E}{v} \frac{\dot  f_2}{f_2} \,.
\end{equation}
Consistency with Eqs. (\ref{pev1})  and  (\ref{pev1theta}) then requires that
\begin{equation}
p=\frac{\int {\mathcal L}_{\rm m} \,  d^3 r +E}{v}\,,
\end{equation}
Taking into account that $p=v E=E_{[\rm prop]} \gamma v$ and Eq. (\ref{traceT}), this in turn implies that 
\begin{equation}
\int {\mathcal L}_{\rm m} d^3 r = -\frac{E}{\gamma^2} =  -\frac{E_{[\rm prop]}}{\gamma} = \int T \, d^3 r\,.
\end{equation}
Hence, the volume average of the on-shell Lagrangian of a particle of fixed mass and structure is equal to the volume average of the trace of its energy-momentum tensor, independently of the particle structure and composition.

\subsection{Perfect fluids in FLRW spacetimes}

Here, we shall derive the dynamics of moving localized particles with fixed proper mass and structure in an FLRW assuming that they are part of a homogeneous perfect fluid, but now in the context of NMC gravity.

\subsubsection{Linear momentum constraints: dust}

In the case of dust, a perfect fluid with $\mathcal{P}_{\rm dust}=0$, the $x$-component of Eq. (\ref{UevNMC}) may be written as
\be
\dot U^1 U^0+ 2 \Gamma^1_{1 0} U^0 U^1= \frac{\mathcal{L}_{\rm dust} }{\rho_{\rm dust}} \frac{\dot f_2}{f_2} U^0 U^1 \,.
\ee
Multiplying this equation by $E_{[\rm prop]} a/U^0$, taking into account that $U^1=\gamma v/a$ and that $\Gamma^0_{1 1}=H$, one obtains 
\be
\dot p + H p = \frac{\mathcal{L}_{\rm dust} }{\rho_{\rm dust}} \frac{\dot f_2}{f_2} p \,. \label{pevNMC}
\ee
Consistency with Eqs. (\ref{pev1})  and  (\ref{pev1theta}) requires that
\begin{equation}
	\mathcal{L}_{\rm dust}=\mathcal -\rho_{\rm dust}= \mathcal{T}_{\rm dust} \,. \label{dustlag}
\end{equation}

\subsubsection{Energy constraints: dust}

The time component of Eq. (\ref{UevNMCdust}) is given by
\be
\dot U^0 U^0+  \Gamma^0_{0 0} U^0 U^0 + \Gamma^0_{1 1} U^1 U^1=\frac{\mathcal{L}_{\rm dust} }{\mathcal \rho_{\rm dust}} \frac{\dot f_2}{f_2} (g^{00}+U^0 U^0)  \,,
\ee
Multiplying this equation by $E_{[\rm prop]} a/U^0$, taking into account that  $g^{00}=-1/a^2$, $U^0=\gamma/a$,  $U^1=\gamma v/a$, $-1+\gamma^2=v^2\gamma^2$, and that $\Gamma^0_{0 0}=\Gamma^0_{1 1}=H$, one obtains
\begin{equation}
{\dot E}+ {H} v^2 E=\frac{\mathcal{L}_{\rm dust} }{\mathcal \rho_{\rm dust}} \frac{\dot f_2}{f_2} E v^2\,. \label{EevNMC}
\end{equation}
Dividing Eq. (\ref{Eev1}) by $v$, taking into account Eqs. (\ref{peq}) and (\ref{dotp}), one again arrives at Eq. (\ref{pevNMC}) for the evolution of linear momentum. 

\subsubsection{Energy constraints: homogeneous and isotropic fluid}

Consider a homogeneous and isotropic perfect fluid (at rest in the comoving frame, so that $U^i=0$) made up of localized particles of fixed mass and structure all with the same speed $v$. The time component of Eq. (\ref{Tncons}), is given by
\be
{\dot \rho_{\rm f}} +  3 H (\rho_{\rm f} +\mathcal P_{\rm f}) =  -({\mathcal L}_{\rm f}  + \rho_{\rm f}) \frac{\dot f_2}{f_2}\,,\label{Tcons2}
\ee
where ${\mathcal L}_{\rm f}$, $\rho_f$ and $\mathcal{P}_{\rm f}$ are the on-shell Lagrangian, proper energy density and proper pressure of the fluid, respectively. If the number of particles is conserved then Eq. (\ref{rhoEV}) is satisfied. On the other hand, if the perfect fluid is an ideal gas then its proper pressure is given by Eq. (\ref{Pideal}): $\mathcal P_{\rm f}=\rho_{\rm f} v^2/3$. Substituting the conditions given in Eqs. (\ref{rhoEV}) and (\ref{Pideal}) into Eq. (\ref{Tcons2}) and multiplying it by $a^3/N$, one obtains
\begin{equation}
{\dot E}+ {H}  v^2 E=  -\left(\frac{{\mathcal L}_{\rm f}}{\rho_{\rm f}}  + 1\right) \frac{\dot f_2}{f_2}E\,. \label{Eevf}
\end{equation}
As in Sec. \ref{sec:emGRF}, this homogeneous and isotropic perfect fluid can be pictured as the combination of six equal density dust fluid components moving in the positive/negative $x$, $y$, and $z$ directions. Therefore, in the proper frame of the resulting perfect fluid, the evolution of particle energy and linear momentum of each of its dust components and of the total combined fluid must be the same, \textit{i.e.} Eqs. (\ref{EevNMC}) and \eqref{Eevf} must result in the same equation of motion. This implies that
\begin{equation}
-\frac{\mathcal{L}_{\rm dust} }{\rho_\text{dust}}v^2=\frac{\mathcal{L}_{\rm f} }{\rho_\text{f}}+1 \,. \label{finald}
\end{equation}
We can therefore write the on-shell Lagrangian of the perfect fluid as
\begin{align}
	\mathcal{L}_\text{f} &= -\rho_\text{f}\left(\frac{\mathcal{L}_\text{dust}}{\rho_{\rm dust}}v^2+1\right)=\rho_\text{f}\left(v^2-1\right)  \nonumber \\
	\Rightarrow \mathcal{L}_\text{f} &= 3\mathcal{P}_{\rm f}-\rho_\text{f} = \mathcal{T}_{\rm f} \,,\label{lagperf}
\end{align}
where we have taken into account Eqs. \eqref{Pideal} and \eqref{dustlag}. Naturally, in the case of dust ($v=0$) Eq. \eqref{lagperf} again implies that $\mathcal{L}_\text{dust}=\mathcal{T}_{\rm dust}=-\rho_\text{dust}$.

\section{Discussion and conclusions}
\label{sec:conc}

The recent increase in research on NMC gravity has brought further attention to the importance of using the correct form of the on-shell Lagrangian of the matter fields in order to provide an accurate description of the corresponding physical implications.

In this work we demonstrated that the equation for the evolution of the linear momentum of a localized particle with fixed mass and structure in a broad class of NMC gravity theories may be univocally determined independently of any specific assumptions about the form of the on-shell Lagrangian of the matter fields. We have further shown that consistency between energy and linear-momentum evolution requires that their volume-averaged on-shell Lagrangian and energy-momentum tensor trace coincide. Despite being derived in the context of NMC theories of gravity, this is a general result, valid regardless of the type of coupling with gravity or other fields, as long as gravity plays a negligible role on the particle structure. We emphasize that the equality $\mathcal L_{\rm on-shell}=T$ is only true on average and is not expected to hold everywhere inside a particle. 

%For example, in the case of the proton although the volume average of the proper pressure is zero, as required by the von Laue condition, there is experimental  evidence for a strong repulsive (positive) pressure at distances up to 0.6 femtometers from the center of the proton and a (negative) confining pressure at larger distances \cite{2018Natur.557..396B}.

We also found that consistency between the evolution of the energy and linear momentum in the context of theories of gravity with an NMC coupling to the matter fields requires that the condition $\mathcal L_{\rm on-shell}=T$ is again satisfied when applied to an ideal gas. In the derivation of this result the crucial assumption is that the fluid can be described by the ideal gas equation of state --- no assumptions have been made regarding the role of gravity on the structure of the particles in this case.  The fact that $\mathcal L_{\rm on-shell}=T$  is required for consistency, gives no margin to other possibilities for the on-shell Lagrangian of an ideal gas. Notice that this result is not in contradiction with the findings of Refs. \cite{Harko2010a,Minazzoli2012}, according to which the on-shell Lagrangian of a fluid with: 1. a conserved number of particles and 2. an off-shell Lagrangian dependent solely on the particle number number density is ${\mathcal L}_{\rm on-shell}=-\rho$, since the second condition does not apply to an ideal gas.
 
%%%%%%%%%%%%%%%%%%%%%%%%%%%%%%%%%%%%%%%%%%%%%%%%%%%%%
\begin{acknowledgments}

Funding of this work has been provided by FCT through national funds (PTDC/FIS-PAR/31938/2017) and by FEDER—Fundo Europeu de Desenvolvimento Regional through COMPETE2020 - Programme for Competitiveness and Internationalisation (POCI-01-0145-FEDER-031938), and through the research grants UIDB/04434/2020 and UIDP/04434/2020. R.P.L.A. was supported by the Funda{\c c}\~ao para a Ci\^encia e Tecnologia (FCT, Portugal) Grant No. SFRH/BD/132546/2017.

\end{acknowledgments}
%%%%%%%%%%%%%%%%%%%%%%%%%%%%%%%%%%%%%%%%%%%%%%%%%%%%%%%%%%

\bibliography{Lagrangian}

\begin{thebibliography}{42}%
\makeatletter
\providecommand \@ifxundefined [1]{%
 \@ifx{#1\undefined}
}%
\providecommand \@ifnum [1]{%
 \ifnum #1\expandafter \@firstoftwo
 \else \expandafter \@secondoftwo
 \fi
}%
\providecommand \@ifx [1]{%
 \ifx #1\expandafter \@firstoftwo
 \else \expandafter \@secondoftwo
 \fi
}%
\providecommand \natexlab [1]{#1}%
\providecommand \enquote  [1]{``#1''}%
\providecommand \bibnamefont  [1]{#1}%
\providecommand \bibfnamefont [1]{#1}%
\providecommand \citenamefont [1]{#1}%
\providecommand \href@noop [0]{\@secondoftwo}%
\providecommand \href [0]{\begingroup \@sanitize@url \@href}%
\providecommand \@href[1]{\@@startlink{#1}\@@href}%
\providecommand \@@href[1]{\endgroup#1\@@endlink}%
\providecommand \@sanitize@url [0]{\catcode `\\12\catcode `\$12\catcode
  `\&12\catcode `\#12\catcode `\^12\catcode `\_12\catcode `\%12\relax}%
\providecommand \@@startlink[1]{}%
\providecommand \@@endlink[0]{}%
\providecommand \url  [0]{\begingroup\@sanitize@url \@url }%
\providecommand \@url [1]{\endgroup\@href {#1}{\urlprefix }}%
\providecommand \urlprefix  [0]{URL }%
\providecommand \Eprint [0]{\href }%
\providecommand \doibase [0]{https://doi.org/}%
\providecommand \selectlanguage [0]{\@gobble}%
\providecommand \bibinfo  [0]{\@secondoftwo}%
\providecommand \bibfield  [0]{\@secondoftwo}%
\providecommand \translation [1]{[#1]}%
\providecommand \BibitemOpen [0]{}%
\providecommand \bibitemStop [0]{}%
\providecommand \bibitemNoStop [0]{.\EOS\space}%
\providecommand \EOS [0]{\spacefactor3000\relax}%
\providecommand \BibitemShut  [1]{\csname bibitem#1\endcsname}%
\let\auto@bib@innerbib\@empty
%</preamble>
\bibitem [{\citenamefont {Schutz}(1970)}]{1970PhRvD...2.2762S}%
  \BibitemOpen
  \bibfield  {author} {\bibinfo {author} {\bibfnamefont {B.~F.}\ \bibnamefont
  {Schutz}},\ }\bibfield  {title} {\bibinfo {title} {{Perfect Fluids in General
  Relativity: Velocity Potentials and a Variational Principle}},\ }\href
  {https://doi.org/10.1103/PhysRevD.2.2762} {\bibfield  {journal} {\bibinfo
  {journal} {Phys. Rev. D}\ }\textbf {\bibinfo {volume} {2}},\ \bibinfo {pages}
  {2762} (\bibinfo {year} {1970})}\BibitemShut {NoStop}%
\bibitem [{\citenamefont {Schutz}\ and\ \citenamefont
  {Sorkin}(1977)}]{Schutz1977}%
  \BibitemOpen
  \bibfield  {author} {\bibinfo {author} {\bibfnamefont {B.~F.}\ \bibnamefont
  {Schutz}}\ and\ \bibinfo {author} {\bibfnamefont {R.}~\bibnamefont
  {Sorkin}},\ }\bibfield  {title} {\bibinfo {title} {{Variational aspects of
  relativistic field theories, with application to perfect fluids}},\ }\href
  {https://doi.org/10.1016/0003-4916(77)90200-7} {\bibfield  {journal}
  {\bibinfo  {journal} {Ann. Phys. (N. Y).}\ }\textbf {\bibinfo {volume}
  {107}},\ \bibinfo {pages} {1} (\bibinfo {year} {1977})}\BibitemShut {NoStop}%
\bibitem [{\citenamefont {Brown}(1993)}]{Brown1993a}%
  \BibitemOpen
  \bibfield  {author} {\bibinfo {author} {\bibfnamefont {J.~D.}\ \bibnamefont
  {Brown}},\ }\bibfield  {title} {\bibinfo {title} {{Action functionals for
  relativistic perfect fluids}},\ }\href
  {https://doi.org/10.1088/0264-9381/10/8/017} {\bibfield  {journal} {\bibinfo
  {journal} {Class. Quantum Gravity}\ }\textbf {\bibinfo {volume} {10}},\
  \bibinfo {pages} {1579} (\bibinfo {year} {1993})},\ \Eprint
  {https://arxiv.org/abs/9304026} {9304026 [gr-qc]} \BibitemShut {NoStop}%
\bibitem [{\citenamefont {Bertolami}\ \emph {et~al.}(2007)\citenamefont
  {Bertolami}, \citenamefont {B{\"{o}}hmer}, \citenamefont {Harko},\ and\
  \citenamefont {Lobo}}]{Bertolami2007}%
  \BibitemOpen
  \bibfield  {author} {\bibinfo {author} {\bibfnamefont {O.}~\bibnamefont
  {Bertolami}}, \bibinfo {author} {\bibfnamefont {C.~G.}\ \bibnamefont
  {B{\"{o}}hmer}}, \bibinfo {author} {\bibfnamefont {T.}~\bibnamefont
  {Harko}},\ and\ \bibinfo {author} {\bibfnamefont {F.~S.~N.}\ \bibnamefont
  {Lobo}},\ }\bibfield  {title} {\bibinfo {title} {{Extra force in $f(R)$
  modified theories of gravity}},\ }\href
  {https://doi.org/10.1103/PhysRevD.75.104016} {\bibfield  {journal} {\bibinfo
  {journal} {Phys. Rev. D}\ }\textbf {\bibinfo {volume} {75}},\ \bibinfo
  {pages} {104016} (\bibinfo {year} {2007})}\BibitemShut {NoStop}%
\bibitem [{\citenamefont {Bertolami}\ \emph {et~al.}(2008)\citenamefont
  {Bertolami}, \citenamefont {Lobo},\ and\ \citenamefont
  {P{\'{a}}ramos}}]{Bertolami:2008ab}%
  \BibitemOpen
  \bibfield  {author} {\bibinfo {author} {\bibfnamefont {O.}~\bibnamefont
  {Bertolami}}, \bibinfo {author} {\bibfnamefont {F.~S.~N.}\ \bibnamefont
  {Lobo}},\ and\ \bibinfo {author} {\bibfnamefont {J.}~\bibnamefont
  {P{\'{a}}ramos}},\ }\bibfield  {title} {\bibinfo {title} {{Nonminimal
  coupling of perfect fluids to curvature}},\ }\href
  {https://doi.org/10.1103/PhysRevD.78.064036} {\bibfield  {journal} {\bibinfo
  {journal} {Phys. Rev. D}\ }\textbf {\bibinfo {volume} {78}},\ \bibinfo
  {pages} {064036} (\bibinfo {year} {2008})}\BibitemShut {NoStop}%
\bibitem [{\citenamefont {Harko}\ \emph {et~al.}(2011)\citenamefont {Harko},
  \citenamefont {Lobo}, \citenamefont {Nojiri},\ and\ \citenamefont
  {Odintsov}}]{Harko:2011kv}%
  \BibitemOpen
  \bibfield  {author} {\bibinfo {author} {\bibfnamefont {T.}~\bibnamefont
  {Harko}}, \bibinfo {author} {\bibfnamefont {F.~S.~N.}\ \bibnamefont {Lobo}},
  \bibinfo {author} {\bibfnamefont {S.}~\bibnamefont {Nojiri}},\ and\ \bibinfo
  {author} {\bibfnamefont {S.~D.}\ \bibnamefont {Odintsov}},\ }\bibfield
  {title} {\bibinfo {title} {{$f(R,T)$ gravity}},\ }\href
  {https://doi.org/10.1103/PhysRevD.84.024020} {\bibfield  {journal} {\bibinfo
  {journal} {Phys. Rev. D}\ }\textbf {\bibinfo {volume} {84}},\ \bibinfo
  {pages} {024020} (\bibinfo {year} {2011})}\BibitemShut {NoStop}%
\bibitem [{\citenamefont {Bahamonde}(2018)}]{2018EPJC...78..326B}%
  \BibitemOpen
  \bibfield  {author} {\bibinfo {author} {\bibfnamefont {S.}~\bibnamefont
  {Bahamonde}},\ }\bibfield  {title} {\bibinfo {title} {{Generalised
  nonminimally gravity-matter coupled theory}},\ }\href
  {https://doi.org/10.1140/epjc/s10052-018-5793-1} {\bibfield  {journal}
  {\bibinfo  {journal} {Eur. Phys. J. C}\ }\textbf {\bibinfo {volume} {78}},\
  \bibinfo {pages} {326} (\bibinfo {year} {2018})}\BibitemShut {NoStop}%
\bibitem [{\citenamefont {Barrientos}\ \emph {et~al.}(2018)\citenamefont
  {Barrientos}, \citenamefont {Lobo}, \citenamefont {Mendoza}, \citenamefont
  {Olmo},\ and\ \citenamefont {Rubiera-Garcia}}]{2018PhRvD..97j4041B}%
  \BibitemOpen
  \bibfield  {author} {\bibinfo {author} {\bibfnamefont {E.}~\bibnamefont
  {Barrientos}}, \bibinfo {author} {\bibfnamefont {F.~S.}\ \bibnamefont
  {Lobo}}, \bibinfo {author} {\bibfnamefont {S.}~\bibnamefont {Mendoza}},
  \bibinfo {author} {\bibfnamefont {G.~J.}\ \bibnamefont {Olmo}},\ and\
  \bibinfo {author} {\bibfnamefont {D.}~\bibnamefont {Rubiera-Garcia}},\
  }\bibfield  {title} {\bibinfo {title} {{Metric-affine $f(R,T)$ theories of
  gravity and their applications}},\ }\href
  {https://doi.org/10.1103/PhysRevD.97.104041} {\bibfield  {journal} {\bibinfo
  {journal} {Phys. Rev. D}\ }\textbf {\bibinfo {volume} {97}},\ \bibinfo
  {pages} {104041} (\bibinfo {year} {2018})}\BibitemShut {NoStop}%
\bibitem [{\citenamefont {Minazzoli}(2018)}]{2018PhRvD..98l4020M}%
  \BibitemOpen
  \bibfield  {author} {\bibinfo {author} {\bibfnamefont {O.}~\bibnamefont
  {Minazzoli}},\ }\bibfield  {title} {\bibinfo {title} {{Rethinking the link
  between matter and geometry}},\ }\href
  {https://doi.org/10.1103/PhysRevD.98.124020} {\bibfield  {journal} {\bibinfo
  {journal} {Phys. Rev. D}\ }\textbf {\bibinfo {volume} {98}},\ \bibinfo
  {pages} {124020} (\bibinfo {year} {2018})}\BibitemShut {NoStop}%
\bibitem [{\citenamefont {Fox}(2019)}]{2019PhRvD..99l4027F}%
  \BibitemOpen
  \bibfield  {author} {\bibinfo {author} {\bibfnamefont {M.~S.}\ \bibnamefont
  {Fox}},\ }\bibfield  {title} {\bibinfo {title} {{Palatini
  $f\mathbf{(}\mathcal{R},{\mathcal{L}}_{m},{\mathcal{R}}_{\ensuremath{\mu}\ensuremath{\nu}}{T}^{\ensuremath{\mu}\ensuremath{\nu}}\mathbf{)}$
  gravity and its Born-Infeld semblance}},\ }\href
  {https://doi.org/10.1103/PhysRevD.99.124027} {\bibfield  {journal} {\bibinfo
  {journal} {Phys. Rev. D}\ }\textbf {\bibinfo {volume} {99}},\ \bibinfo
  {pages} {124027} (\bibinfo {year} {2019})}\BibitemShut {NoStop}%
\bibitem [{\citenamefont {Harko}(2010)}]{Harko2010a}%
  \BibitemOpen
  \bibfield  {author} {\bibinfo {author} {\bibfnamefont {T.}~\bibnamefont
  {Harko}},\ }\bibfield  {title} {\bibinfo {title} {{The matter Lagrangian and
  the energy-momentum tensor in modified gravity with nonminimal coupling
  between matter and geometry}},\ }\href
  {https://doi.org/10.1103/PhysRevD.81.044021} {\bibfield  {journal} {\bibinfo
  {journal} {Phys. Rev. D}\ }\textbf {\bibinfo {volume} {81}},\ \bibinfo
  {pages} {44021} (\bibinfo {year} {2010})}\BibitemShut {NoStop}%
\bibitem [{\citenamefont {Minazzoli}\ and\ \citenamefont
  {Harko}(2012)}]{Minazzoli2012}%
  \BibitemOpen
  \bibfield  {author} {\bibinfo {author} {\bibfnamefont {O.}~\bibnamefont
  {Minazzoli}}\ and\ \bibinfo {author} {\bibfnamefont {T.}~\bibnamefont
  {Harko}},\ }\bibfield  {title} {\bibinfo {title} {{New derivation of the
  Lagrangian of a perfect fluid with a barotropic equation of state}},\ }\href
  {https://doi.org/10.1103/PhysRevD.86.087502} {\bibfield  {journal} {\bibinfo
  {journal} {Phys. Rev. D}\ }\textbf {\bibinfo {volume} {86}},\ \bibinfo
  {pages} {087502} (\bibinfo {year} {2012})}\BibitemShut {NoStop}%
\bibitem [{\citenamefont {Minazzoli}(2013)}]{Minazzoli2013}%
  \BibitemOpen
  \bibfield  {author} {\bibinfo {author} {\bibfnamefont {O.}~\bibnamefont
  {Minazzoli}},\ }\bibfield  {title} {\bibinfo {title} {{Conservation laws in
  theories with universal gravity/matter coupling}},\ }\href
  {https://doi.org/10.1103/PhysRevD.88.027506} {\bibfield  {journal} {\bibinfo
  {journal} {Phys. Rev. D}\ }\textbf {\bibinfo {volume} {88}},\ \bibinfo
  {pages} {027506} (\bibinfo {year} {2013})}\BibitemShut {NoStop}%
\bibitem [{\citenamefont {Ferreira}\ \emph {et~al.}(2020)\citenamefont
  {Ferreira}, \citenamefont {Avelino},\ and\ \citenamefont
  {Azevedo}}]{Ferreira:2020fma}%
  \BibitemOpen
  \bibfield  {author} {\bibinfo {author} {\bibfnamefont {V.~M.~C.}\
  \bibnamefont {Ferreira}}, \bibinfo {author} {\bibfnamefont {P.~P.}\
  \bibnamefont {Avelino}},\ and\ \bibinfo {author} {\bibfnamefont {R.~P.~L.}\
  \bibnamefont {Azevedo}},\ }\bibfield  {title} {\bibinfo {title} {{Lagrangian
  description of cosmic fluids: Mapping dark energy into unified dark
  energy}},\ }\href {https://doi.org/10.1103/PhysRevD.102.063525} {\bibfield
  {journal} {\bibinfo  {journal} {Phys. Rev. D}\ }\textbf {\bibinfo {volume}
  {102}},\ \bibinfo {pages} {063525} (\bibinfo {year} {2020})}\BibitemShut
  {NoStop}%
\bibitem [{\citenamefont {Arruga}\ \emph {et~al.}(2021)\citenamefont {Arruga},
  \citenamefont {Rousselle},\ and\ \citenamefont {Minazzoli}}]{Arruga2021}%
  \BibitemOpen
  \bibfield  {author} {\bibinfo {author} {\bibfnamefont {D.}~\bibnamefont
  {Arruga}}, \bibinfo {author} {\bibfnamefont {O.}~\bibnamefont {Rousselle}},\
  and\ \bibinfo {author} {\bibfnamefont {O.}~\bibnamefont {Minazzoli}},\
  }\bibfield  {title} {\bibinfo {title} {{Compact objects in entangled
  relativity}},\ }\href {https://doi.org/10.1103/PhysRevD.103.024034}
  {\bibfield  {journal} {\bibinfo  {journal} {Phys. Rev. D}\ }\textbf {\bibinfo
  {volume} {103}},\ \bibinfo {pages} {024034} (\bibinfo {year} {2021})},\
  \Eprint {https://arxiv.org/abs/2011.14629} {arXiv:2011.14629} \BibitemShut
  {NoStop}%
\bibitem [{\citenamefont {Avelino}\ and\ \citenamefont
  {Azevedo}(2018)}]{Avelino:2018rsb}%
  \BibitemOpen
  \bibfield  {author} {\bibinfo {author} {\bibfnamefont {P.~P.}\ \bibnamefont
  {Avelino}}\ and\ \bibinfo {author} {\bibfnamefont {R.~P.~L.}\ \bibnamefont
  {Azevedo}},\ }\bibfield  {title} {\bibinfo {title} {{Perfect fluid Lagrangian
  and its cosmological implications in theories of gravity with nonminimally
  coupled matter fields}},\ }\href {https://doi.org/10.1103/PhysRevD.97.064018}
  {\bibfield  {journal} {\bibinfo  {journal} {Phys. Rev. D}\ }\textbf {\bibinfo
  {volume} {97}},\ \bibinfo {pages} {064018} (\bibinfo {year}
  {2018})}\BibitemShut {NoStop}%
\bibitem [{\citenamefont {Avelino}\ and\ \citenamefont
  {Sousa}(2018)}]{Avelino:2018qgt}%
  \BibitemOpen
  \bibfield  {author} {\bibinfo {author} {\bibfnamefont {P.~P.}\ \bibnamefont
  {Avelino}}\ and\ \bibinfo {author} {\bibfnamefont {L.}~\bibnamefont
  {Sousa}},\ }\bibfield  {title} {\bibinfo {title} {{Matter Lagrangian of
  particles and fluids}},\ }\href {https://doi.org/10.1103/PhysRevD.97.064019}
  {\bibfield  {journal} {\bibinfo  {journal} {Phys. Rev. D}\ }\textbf {\bibinfo
  {volume} {97}},\ \bibinfo {pages} {064019} (\bibinfo {year}
  {2018})}\BibitemShut {NoStop}%
\bibitem [{\citenamefont {Bekenstein}(1982)}]{Bekenstein1982}%
  \BibitemOpen
  \bibfield  {author} {\bibinfo {author} {\bibfnamefont {J.~D.}\ \bibnamefont
  {Bekenstein}},\ }\bibfield  {title} {\bibinfo {title} {{Fine-structure
  constant: Is it really a constant?}},\ }\href
  {https://doi.org/10.1103/PhysRevD.25.1527} {\bibfield  {journal} {\bibinfo
  {journal} {Phys. Rev. D}\ }\textbf {\bibinfo {volume} {25}},\ \bibinfo
  {pages} {1527} (\bibinfo {year} {1982})}\BibitemShut {NoStop}%
\bibitem [{\citenamefont {Sandvik}\ \emph {et~al.}(2002)\citenamefont
  {Sandvik}, \citenamefont {Barrow},\ and\ \citenamefont
  {Magueijo}}]{Sandvik2002}%
  \BibitemOpen
  \bibfield  {author} {\bibinfo {author} {\bibfnamefont {H.~B.}\ \bibnamefont
  {Sandvik}}, \bibinfo {author} {\bibfnamefont {J.~D.}\ \bibnamefont
  {Barrow}},\ and\ \bibinfo {author} {\bibfnamefont {J.}~\bibnamefont
  {Magueijo}},\ }\bibfield  {title} {\bibinfo {title} {{A Simple Cosmology with
  a Varying Fine Structure Constant}},\ }\href
  {https://doi.org/10.1103/PhysRevLett.88.031302} {\bibfield  {journal}
  {\bibinfo  {journal} {Phys. Rev. Lett.}\ }\textbf {\bibinfo {volume} {88}},\
  \bibinfo {pages} {031302} (\bibinfo {year} {2002})}\BibitemShut {NoStop}%
\bibitem [{\citenamefont {Anchordoqui}\ and\ \citenamefont
  {Goldberg}(2003)}]{Anchordoqui2003}%
  \BibitemOpen
  \bibfield  {author} {\bibinfo {author} {\bibfnamefont {L.}~\bibnamefont
  {Anchordoqui}}\ and\ \bibinfo {author} {\bibfnamefont {H.}~\bibnamefont
  {Goldberg}},\ }\bibfield  {title} {\bibinfo {title} {{Time variation of the
  fine structure constant driven by quintessence}},\ }\href
  {https://doi.org/10.1103/PhysRevD.68.083513} {\bibfield  {journal} {\bibinfo
  {journal} {Phys. Rev. D}\ }\textbf {\bibinfo {volume} {68}},\ \bibinfo
  {pages} {083513} (\bibinfo {year} {2003})}\BibitemShut {NoStop}%
\bibitem [{\citenamefont {Copeland}\ \emph {et~al.}(2004)\citenamefont
  {Copeland}, \citenamefont {Nunes},\ and\ \citenamefont
  {Pospelov}}]{Copeland2004}%
  \BibitemOpen
  \bibfield  {author} {\bibinfo {author} {\bibfnamefont {E.~J.}\ \bibnamefont
  {Copeland}}, \bibinfo {author} {\bibfnamefont {N.~J.}\ \bibnamefont
  {Nunes}},\ and\ \bibinfo {author} {\bibfnamefont {M.}~\bibnamefont
  {Pospelov}},\ }\bibfield  {title} {\bibinfo {title} {{Models of quintessence
  coupled to the electromagnetic field and the cosmological evolution of
  alpha}},\ }\href {https://doi.org/10.1103/PhysRevD.69.023501} {\bibfield
  {journal} {\bibinfo  {journal} {Phys. Rev. D}\ }\textbf {\bibinfo {volume}
  {69}},\ \bibinfo {pages} {023501} (\bibinfo {year} {2004})}\BibitemShut
  {NoStop}%
\bibitem [{\citenamefont {Lee}\ \emph {et~al.}(2004)\citenamefont {Lee},
  \citenamefont {Olive},\ and\ \citenamefont {Pospelov}}]{Lee2004}%
  \BibitemOpen
  \bibfield  {author} {\bibinfo {author} {\bibfnamefont {S.}~\bibnamefont
  {Lee}}, \bibinfo {author} {\bibfnamefont {K.~A.}\ \bibnamefont {Olive}},\
  and\ \bibinfo {author} {\bibfnamefont {M.}~\bibnamefont {Pospelov}},\
  }\bibfield  {title} {\bibinfo {title} {{Quintessence models and the
  cosmological evolution of $\alpha$}},\ }\href
  {https://doi.org/10.1103/PhysRevD.70.083503} {\bibfield  {journal} {\bibinfo
  {journal} {Phys. Rev. D}\ }\textbf {\bibinfo {volume} {70}},\ \bibinfo
  {pages} {083503} (\bibinfo {year} {2004})}\BibitemShut {NoStop}%
\bibitem [{\citenamefont {Koivisto}(2005)}]{Koivisto2005}%
  \BibitemOpen
  \bibfield  {author} {\bibinfo {author} {\bibfnamefont {T.}~\bibnamefont
  {Koivisto}},\ }\bibfield  {title} {\bibinfo {title} {{Growth of perturbations
  in dark matter coupled with quintessence}},\ }\href
  {https://doi.org/10.1103/PhysRevD.72.043516} {\bibfield  {journal} {\bibinfo
  {journal} {Phys. Rev. D}\ }\textbf {\bibinfo {volume} {72}},\ \bibinfo
  {pages} {043516} (\bibinfo {year} {2005})}\BibitemShut {NoStop}%
\bibitem [{\citenamefont {Avelino}(2008)}]{Avelino2008}%
  \BibitemOpen
  \bibfield  {author} {\bibinfo {author} {\bibfnamefont {P.~P.}\ \bibnamefont
  {Avelino}},\ }\bibfield  {title} {\bibinfo {title} {{Cosmological evolution
  of $\alpha$ and $\mu$ and the dynamics of dark energy}},\ }\href
  {https://doi.org/10.1103/PhysRevD.78.043516} {\bibfield  {journal} {\bibinfo
  {journal} {Phys. Rev. D}\ }\textbf {\bibinfo {volume} {78}},\ \bibinfo
  {pages} {043516} (\bibinfo {year} {2008})}\BibitemShut {NoStop}%
\bibitem [{\citenamefont {Bettoni}\ \emph {et~al.}(2011)\citenamefont
  {Bettoni}, \citenamefont {Liberati},\ and\ \citenamefont
  {Sindoni}}]{Bettoni2011}%
  \BibitemOpen
  \bibfield  {author} {\bibinfo {author} {\bibfnamefont {D.}~\bibnamefont
  {Bettoni}}, \bibinfo {author} {\bibfnamefont {S.}~\bibnamefont {Liberati}},\
  and\ \bibinfo {author} {\bibfnamefont {L.}~\bibnamefont {Sindoni}},\
  }\bibfield  {title} {\bibinfo {title} {{Extended $\Lambda$CDM: generalized
  non-minimal coupling for dark matter fluids}},\ }\href
  {https://doi.org/10.1088/1475-7516/2011/11/007} {\bibfield  {journal}
  {\bibinfo  {journal} {J. Cosmol. Astropart. Phys.}\ }\textbf {\bibinfo
  {volume} {2011}}\bibinfo  {number} { (11)},\ \bibinfo {pages}
  {007}}\BibitemShut {NoStop}%
\bibitem [{\citenamefont {Ayaita}\ \emph {et~al.}(2012)\citenamefont {Ayaita},
  \citenamefont {Weber},\ and\ \citenamefont {Wetterich}}]{Ayaita2012}%
  \BibitemOpen
\bibfield  {number} {  }\bibfield  {author} {\bibinfo {author} {\bibfnamefont
  {Y.}~\bibnamefont {Ayaita}}, \bibinfo {author} {\bibfnamefont
  {M.}~\bibnamefont {Weber}},\ and\ \bibinfo {author} {\bibfnamefont
  {C.}~\bibnamefont {Wetterich}},\ }\bibfield  {title} {\bibinfo {title}
  {{Structure formation and backreaction in growing neutrino quintessence}},\
  }\href {https://doi.org/10.1103/PhysRevD.85.123010} {\bibfield  {journal}
  {\bibinfo  {journal} {Phys. Rev. D}\ }\textbf {\bibinfo {volume} {85}},\
  \bibinfo {pages} {123010} (\bibinfo {year} {2012})}\BibitemShut {NoStop}%
\bibitem [{\citenamefont {Pourtsidou}\ \emph {et~al.}(2013)\citenamefont
  {Pourtsidou}, \citenamefont {Skordis},\ and\ \citenamefont
  {Copeland}}]{Pourtsidou2013}%
  \BibitemOpen
  \bibfield  {author} {\bibinfo {author} {\bibfnamefont {A.}~\bibnamefont
  {Pourtsidou}}, \bibinfo {author} {\bibfnamefont {C.}~\bibnamefont
  {Skordis}},\ and\ \bibinfo {author} {\bibfnamefont {E.~J.}\ \bibnamefont
  {Copeland}},\ }\bibfield  {title} {\bibinfo {title} {{Models of dark matter
  coupled to dark energy}},\ }\href
  {https://doi.org/10.1103/PhysRevD.88.083505} {\bibfield  {journal} {\bibinfo
  {journal} {Phys. Rev. D}\ }\textbf {\bibinfo {volume} {88}},\ \bibinfo
  {pages} {083505} (\bibinfo {year} {2013})}\BibitemShut {NoStop}%
\bibitem [{\citenamefont {B{\"{o}}hmer}\ \emph
  {et~al.}(2015{\natexlab{a}})\citenamefont {B{\"{o}}hmer}, \citenamefont
  {Tamanini},\ and\ \citenamefont {Wright}}]{Bohmer2015}%
  \BibitemOpen
  \bibfield  {author} {\bibinfo {author} {\bibfnamefont {C.~G.}\ \bibnamefont
  {B{\"{o}}hmer}}, \bibinfo {author} {\bibfnamefont {N.}~\bibnamefont
  {Tamanini}},\ and\ \bibinfo {author} {\bibfnamefont {M.}~\bibnamefont
  {Wright}},\ }\bibfield  {title} {\bibinfo {title} {{Interacting quintessence
  from a variational approach. II. Derivative couplings}},\ }\href
  {https://doi.org/10.1103/PhysRevD.91.123003} {\bibfield  {journal} {\bibinfo
  {journal} {Phys. Rev. D}\ }\textbf {\bibinfo {volume} {91}},\ \bibinfo
  {pages} {123003} (\bibinfo {year} {2015}{\natexlab{a}})}\BibitemShut
  {NoStop}%
\bibitem [{\citenamefont {B{\"{o}}hmer}\ \emph
  {et~al.}(2015{\natexlab{b}})\citenamefont {B{\"{o}}hmer}, \citenamefont
  {Tamanini},\ and\ \citenamefont {Wright}}]{Bohmer2015a}%
  \BibitemOpen
  \bibfield  {author} {\bibinfo {author} {\bibfnamefont {C.~G.}\ \bibnamefont
  {B{\"{o}}hmer}}, \bibinfo {author} {\bibfnamefont {N.}~\bibnamefont
  {Tamanini}},\ and\ \bibinfo {author} {\bibfnamefont {M.}~\bibnamefont
  {Wright}},\ }\bibfield  {title} {\bibinfo {title} {{Einstein static universe
  in scalar-fluid theories}},\ }\href
  {https://doi.org/10.1103/PhysRevD.92.124067} {\bibfield  {journal} {\bibinfo
  {journal} {Phys. Rev. D}\ }\textbf {\bibinfo {volume} {92}},\ \bibinfo
  {pages} {124067} (\bibinfo {year} {2015}{\natexlab{b}})}\BibitemShut
  {NoStop}%
\bibitem [{\citenamefont {Bettoni}\ and\ \citenamefont
  {Liberati}(2015)}]{Bettoni2015}%
  \BibitemOpen
  \bibfield  {author} {\bibinfo {author} {\bibfnamefont {D.}~\bibnamefont
  {Bettoni}}\ and\ \bibinfo {author} {\bibfnamefont {S.}~\bibnamefont
  {Liberati}},\ }\bibfield  {title} {\bibinfo {title} {{Dynamics of
  non-minimally coupled perfect fluids}},\ }\href
  {https://doi.org/10.1088/1475-7516/2015/08/023} {\bibfield  {journal}
  {\bibinfo  {journal} {J. Cosmol. Astropart. Phys.}\ }\textbf {\bibinfo
  {volume} {2015}}\bibinfo  {number} { (08)},\ \bibinfo {pages}
  {023}}\BibitemShut {NoStop}%
\bibitem [{\citenamefont {Koivisto}\ \emph {et~al.}(2015)\citenamefont
  {Koivisto}, \citenamefont {Saridakis},\ and\ \citenamefont
  {Tamanini}}]{Koivisto2015}%
  \BibitemOpen
\bibfield  {number} {  }\bibfield  {author} {\bibinfo {author} {\bibfnamefont
  {T.~S.}\ \bibnamefont {Koivisto}}, \bibinfo {author} {\bibfnamefont {E.~N.}\
  \bibnamefont {Saridakis}},\ and\ \bibinfo {author} {\bibfnamefont
  {N.}~\bibnamefont {Tamanini}},\ }\bibfield  {title} {\bibinfo {title}
  {{Scalar-fluid theories: cosmological perturbations and large-scale
  structure}},\ }\href {https://doi.org/10.1088/1475-7516/2015/09/047}
  {\bibfield  {journal} {\bibinfo  {journal} {J. Cosmol. Astropart. Phys.}\
  }\textbf {\bibinfo {volume} {2015}}\bibinfo  {number} { (09)},\ \bibinfo
  {pages} {047}}\BibitemShut {NoStop}%
\bibitem [{\citenamefont {Brax}\ and\ \citenamefont
  {Tamanini}(2016)}]{Brax2016}%
  \BibitemOpen
\bibfield  {number} {  }\bibfield  {author} {\bibinfo {author} {\bibfnamefont
  {P.}~\bibnamefont {Brax}}\ and\ \bibinfo {author} {\bibfnamefont
  {N.}~\bibnamefont {Tamanini}},\ }\bibfield  {title} {\bibinfo {title}
  {{Extended chameleon models}},\ }\href
  {https://doi.org/10.1103/PhysRevD.93.103502} {\bibfield  {journal} {\bibinfo
  {journal} {Phys. Rev. D}\ }\textbf {\bibinfo {volume} {93}},\ \bibinfo
  {pages} {103502} (\bibinfo {year} {2016})}\BibitemShut {NoStop}%
\bibitem [{\citenamefont {Tamanini}\ and\ \citenamefont
  {Wright}(2016)}]{Tamanini2016}%
  \BibitemOpen
  \bibfield  {author} {\bibinfo {author} {\bibfnamefont {N.}~\bibnamefont
  {Tamanini}}\ and\ \bibinfo {author} {\bibfnamefont {M.}~\bibnamefont
  {Wright}},\ }\bibfield  {title} {\bibinfo {title} {{Cosmological dynamics of
  extended chameleons}},\ }\href
  {https://doi.org/10.1088/1475-7516/2016/04/032} {\bibfield  {journal}
  {\bibinfo  {journal} {J. Cosmol. Astropart. Phys.}\ }\textbf {\bibinfo
  {volume} {2016}}\bibinfo  {number} { (04)},\ \bibinfo {pages}
  {032}}\BibitemShut {NoStop}%
\bibitem [{\citenamefont {Dutta}\ \emph {et~al.}(2017)\citenamefont {Dutta},
  \citenamefont {Khyllep},\ and\ \citenamefont {Tamanini}}]{Dutta2017}%
  \BibitemOpen
\bibfield  {number} {  }\bibfield  {author} {\bibinfo {author} {\bibfnamefont
  {J.}~\bibnamefont {Dutta}}, \bibinfo {author} {\bibfnamefont
  {W.}~\bibnamefont {Khyllep}},\ and\ \bibinfo {author} {\bibfnamefont
  {N.}~\bibnamefont {Tamanini}},\ }\bibfield  {title} {\bibinfo {title}
  {{Scalar-fluid interacting dark energy: Cosmological dynamics beyond the
  exponential potential}},\ }\href {https://doi.org/10.1103/PhysRevD.95.023515}
  {\bibfield  {journal} {\bibinfo  {journal} {Phys. Rev. D}\ }\textbf {\bibinfo
  {volume} {95}},\ \bibinfo {pages} {023515} (\bibinfo {year}
  {2017})}\BibitemShut {NoStop}%
\bibitem [{\citenamefont {Barros}(2019)}]{Barros2019}%
  \BibitemOpen
  \bibfield  {author} {\bibinfo {author} {\bibfnamefont {B.~J.}\ \bibnamefont
  {Barros}},\ }\bibfield  {title} {\bibinfo {title} {{Kinetically coupled dark
  energy}},\ }\href {https://doi.org/10.1103/PhysRevD.99.064051} {\bibfield
  {journal} {\bibinfo  {journal} {Phys. Rev. D}\ }\textbf {\bibinfo {volume}
  {99}},\ \bibinfo {pages} {064051} (\bibinfo {year} {2019})}\BibitemShut
  {NoStop}%
\bibitem [{\citenamefont {Kase}\ and\ \citenamefont
  {Tsujikawa}(2020)}]{Kase2020a}%
  \BibitemOpen
  \bibfield  {author} {\bibinfo {author} {\bibfnamefont {R.}~\bibnamefont
  {Kase}}\ and\ \bibinfo {author} {\bibfnamefont {S.}~\bibnamefont
  {Tsujikawa}},\ }\bibfield  {title} {\bibinfo {title} {{Scalar-field dark
  energy nonminimally and kinetically coupled to dark matter}},\ }\href
  {https://doi.org/10.1103/PhysRevD.101.063511} {\bibfield  {journal} {\bibinfo
   {journal} {Phys. Rev. D}\ }\textbf {\bibinfo {volume} {101}},\ \bibinfo
  {pages} {063511} (\bibinfo {year} {2020})}\BibitemShut {NoStop}%
\bibitem [{\citenamefont {Azevedo}\ and\ \citenamefont
  {Avelino}(2018)}]{Azevedo:2018nvi}%
  \BibitemOpen
  \bibfield  {author} {\bibinfo {author} {\bibfnamefont {R.~P.~L.}\
  \bibnamefont {Azevedo}}\ and\ \bibinfo {author} {\bibfnamefont {P.~P.}\
  \bibnamefont {Avelino}},\ }\bibfield  {title} {\bibinfo {title} {{Big-bang
  nucleosynthesis and cosmic microwave background constraints on nonminimally
  coupled theories of gravity}},\ }\href
  {https://doi.org/10.1103/PhysRevD.98.064045} {\bibfield  {journal} {\bibinfo
  {journal} {Phys. Rev. D}\ }\textbf {\bibinfo {volume} {98}},\ \bibinfo
  {pages} {064045} (\bibinfo {year} {2018})}\BibitemShut {NoStop}%
\bibitem [{\citenamefont {Fisher}\ and\ \citenamefont
  {Carlson}(2022)}]{Fisher:2021nwm}%
  \BibitemOpen
  \bibfield  {author} {\bibinfo {author} {\bibfnamefont {S.~B.}\ \bibnamefont
  {Fisher}}\ and\ \bibinfo {author} {\bibfnamefont {E.~D.}\ \bibnamefont
  {Carlson}},\ }\bibfield  {title} {\bibinfo {title} {{Nuclear limits on
  nonminimally coupled gravity}},\ }\href
  {https://doi.org/10.1103/PhysRevD.105.024020} {\bibfield  {journal} {\bibinfo
   {journal} {Phys. Rev. D}\ }\textbf {\bibinfo {volume} {105}},\ \bibinfo
  {pages} {024020} (\bibinfo {year} {2022})}\BibitemShut {NoStop}%
\bibitem [{\citenamefont {Azevedo}\ and\ \citenamefont
  {Avelino}(2021)}]{Azevedo2021}%
  \BibitemOpen
  \bibfield  {author} {\bibinfo {author} {\bibfnamefont {R.~P.~L.}\
  \bibnamefont {Azevedo}}\ and\ \bibinfo {author} {\bibfnamefont {P.~P.}\
  \bibnamefont {Avelino}},\ }\bibfield  {title} {\bibinfo {title}
  {{Distance-duality in theories with a nonminimal coupling to gravity}},\
  }\href {https://doi.org/10.1103/PhysRevD.104.084079} {\bibfield  {journal}
  {\bibinfo  {journal} {Phys. Rev. D}\ }\textbf {\bibinfo {volume} {104}},\
  \bibinfo {pages} {084079} (\bibinfo {year} {2021})}\BibitemShut {NoStop}%
\bibitem [{\citenamefont {Avelino}\ and\ \citenamefont
  {Azevedo}(2020)}]{Avelino:2020fek}%
  \BibitemOpen
  \bibfield  {author} {\bibinfo {author} {\bibfnamefont {P.}~\bibnamefont
  {Avelino}}\ and\ \bibinfo {author} {\bibfnamefont {R.}~\bibnamefont
  {Azevedo}},\ }\bibfield  {title} {\bibinfo {title} {{Boltzmann's $H$-theorem,
  entropy and the strength of gravity in theories with a nonminimal coupling
  between matter and geometry}},\ }\href
  {https://doi.org/10.1016/j.physletb.2020.135641} {\bibfield  {journal}
  {\bibinfo  {journal} {Phys. Lett. B}\ }\textbf {\bibinfo {volume} {808}},\
  \bibinfo {pages} {135641} (\bibinfo {year} {2020})}\BibitemShut {NoStop}%
\bibitem [{\citenamefont {Azevedo}\ and\ \citenamefont
  {Avelino}(2020)}]{Azevedo:2019oah}%
  \BibitemOpen
  \bibfield  {author} {\bibinfo {author} {\bibfnamefont {R.~P.~L.}\
  \bibnamefont {Azevedo}}\ and\ \bibinfo {author} {\bibfnamefont {P.~P.}\
  \bibnamefont {Avelino}},\ }\bibfield  {title} {\bibinfo {title} {{Second law
  of thermodynamics in nonminimally coupled gravity}},\ }\href
  {https://doi.org/10.1209/0295-5075/132/30005} {\bibfield  {journal} {\bibinfo
   {journal} {Europhys. Lett.}\ }\textbf {\bibinfo {volume} {132}},\ \bibinfo
  {pages} {30005} (\bibinfo {year} {2020})}\BibitemShut {NoStop}%
\bibitem [{\citenamefont {Laue}(1911)}]{doi:10.1002/andp.19113400808}%
  \BibitemOpen
  \bibfield  {author} {\bibinfo {author} {\bibfnamefont {M.}~\bibnamefont
  {Laue}},\ }\bibfield  {title} {\bibinfo {title} {{Zur Dynamik der
  Relativit{\"{a}}tstheorie}},\ }\href
  {https://doi.org/10.1002/andp.19113400808} {\bibfield  {journal} {\bibinfo
  {journal} {Ann. Phys.}\ }\textbf {\bibinfo {volume} {340}},\ \bibinfo {pages}
  {524} (\bibinfo {year} {1911})}\BibitemShut {NoStop}%
\end{thebibliography}%
 	
 \end{document}